# THE PARAMETRIC GENERALIZED FRACTIONAL NIKIFOROV-UVAROV METHOD AND ITS APPLICATIONS†

**M. Abu-Shady[a],\*, H.M. Fath-Allah[b]**

[a]*Department of Mathematics and Computer Sciences, Faculty of Science, Menoufia University, Egypt*
[b]*Higher Institute of Engineering and Technology, Menoufia, Egypt*
*\*Corresponding Author e-mail: dr.abushady@gmail.com*
Received June 27, 2023; revised July 15, 2023; accepted July 16, 2023

By using generalized fractional derivative, the parametric generalized fractional Nikiforov-Uvarov (NU) method is introduced. The second-order parametric generalized differential equation is exactly solved in the fractional form. The obtained results are applied on the extended Cornell potential, the pesudoharmonic potential, the Mie potential, the Kratzer-Fues potential, the harmonic oscillator potential, the Morse potential, the Woods-Saxon potential, the Hulthen potential, the deformed Rosen-Morse potential and the Pöschl-Teller potential which play an important role in the fields of molecular and hadronic physics. The special of classical cases are obtained from the fractional cases at $\alpha = \beta = 1$ which are agreements with recent works.
**Keywords:** *Nonrelativistic models; Generalized fractional derivative; Molecular physics; Hadronic physics*
**PACS:** 12.39.Jh, 31.15.-p, 02.70.−c

## 1. INTRODUCTION

Many researchers have been interested in the fractional calculus (FC) during the past three decades in the previous and current centuries [1-2]. The importance of FC has significantly increased in a variety of scientific and technical fields and explain the advantages of the (FC) over the other numerical methods because we can get exact solution but numerical method can get approximate solution also, there are a lot of problems in solving partial equations. In fact, there are many papers using symmetry methods to solve differential equations including fractional differential equations such as in Ref. [3]. They study modified Gardner-type equation and its time fractional form. They derived these two equations from Fermi-Pasta-Ulam model, and found that these two equations are related with nonlinear Schrodinger equation. They not only derive these two equations, but also use perturbation analysis to find the connection between them and the Schrodinger equation. Non- integer order differentiation and integration form the basis of FC. Numerous definitions of the fractional differential equations have been put out in the literature. The definitions of Jumarie [4], Riemann-Liouville [5], and Caputo [6] have gotten a lot of attention and are the most suitable for physical conditions. Al-Raeei and El-Daher [7], used a numerical technique to rely on the definition of Riemann-Liouville fractional derivative. Khalil et al. [8] represented a new definition of a fractional derivative, referred to as a conformable fractional derivative (CFD) which follows to basic classical principles. Abdeljawad [9] extended the definition and established the fundamental notions of the CFD. A new concept for the fractional derivative known as the generalized fractional derivative (GFD) was recently proposed by Abu-Shady and Kaabar [10]. Because it offers more features than the previous definitions [4-6,10], where the CFD may be produced as a special case from the GFD, the GFD definition is regarded as a comprehensive type for the fractional derivative.

Solving Schrödinger equation (SE) with the intention of examining a physical system is a fundamental challenge in quantum mechanics and particle physics [11-17]. In Ref. [18]. the conformable fractional of the Nikiforov-Uvarov (CF-NU) method is used to analytically solve the radial Schrödinger equation and the dependent temperature potential is used to obtain the energy eigenvalues, corresponding wave functions, and heavy quarkonium masses like charmonium and bottomonium in a hot QCD medium in the 3D and higher dimensions. In Ref. [19], the trigonometric Rosen-Morse potential is employed to examine the effect of the fraction-order parameter. The N-radial fractional Schrödinger equation has analytical solutions established using the extended Nikiforov-Uvarov method. The energy eigenvalues in the fractional forms and the masses of mesons such as charmonium and bottomonium were also obtained.

Using the generalized fractional NU method, the fractional N-dimensional radial Schrödinger equation (SE) with the Deng-Fan potential is evaluated in Ref. [20] in which the analytical formulas are generated for the energy eigenvalues and corresponding eigenfunctions at three-dimensional space and higher dimensions to study the energy spectra of various molecules. The analytical-exact iteration method with a conformable fractional derivative is used in Ref. [21] in which the radial Schrödinger equation can be solved analytically with the trigonometric Rosen-Morse potential. In Ref. [22], the fractional nonrelativistic potential model is used to explore the dissociation of heavy quarkonium in a hot magnetized medium in which the energy eigenvalues and the radial wave functions are obtained. The generalized fractional analytical iteration method is used as in Ref. [23] to solve the hyper-central Schrödinger

---

equation and studied its applications on the theory baryons with single, double, and triple in the ground state. In addition, the SE can be exactly solved, the system can be fully described as in [24, 25, 26] using the Nikiforov Uvarov method. This method is very good because we get on a good result compared by another methods as in Refs. [22, 23].

The aim of the present work is to generalize the second-order parametric differential equation in the fractional form by using generalized fractional derivative. The special cases are obtained at $\alpha = \beta = 1$. Many applications are introduced such as the extended Cornell potential, the Pesudoharmonic potential, the Mie potential, the Kratzer-Fues potential, the harmonic oscillator potential, the Morse potential, the Woods-Saxon potential, the Hulthen potential, the Deformed Rosen-Morse potential and Pöschl-Teller potential. This work is not considered in the recent works.

This paper is arranged as follows. In Sec. 2, the generalized fractional derivative is briefly introduced. In Sec. 3, the generalized fractional Nikiforov-Uvarov (NU) method is explained. In Sec. 4, Some Applications are obtained. In Sec. 5, the conclusion is written.

## 2. THE GENERALIZED FRACTIONAL DERIVATIVE

A new formula for a fractional derivative called the generalized fractional derivative (GFD) is proposed. The generalized fractional Derivative has been suggested to provide more advantages than other classical Caputo and Riemann–Liouville fractional derivative definitions such that the derivative of two functions, the derivative of the quotient of two function, Rolle's theorem and the mean value theorem which have been satisfied in the GFD and which gives a new direction for simply solving fractional differential equations see Ref. [10]. For a function $Z : (0,\infty) \to R$, the generalized fractional derivative of order $0 < \alpha \leq 1$ of $Z(t)$ at $> 0$ is defined as

$$D^{GFD} Z(t) = \lim_{\varepsilon \to 0} \frac{Z\left(t + \frac{\Gamma(\beta)}{\Gamma(\beta-\alpha+1)} \varepsilon t^{1-\alpha}\right) - Z(t)}{\varepsilon}; \beta > -1, \beta \in R^+ \quad (1)$$

The properties of the generalized fractional derivative are,

I.  $D^\alpha[Z_{nl}(t)] = k_1 \, t^{1-\alpha} \, \dot{Z}_{nl}(t),$  (2)
II. $D^\alpha[D^\alpha Z(t)] = k_1^2 \, [(1-\alpha) \, t^{1-2\alpha} \, \dot{Z}_{nl}(t) + t^{2-2\alpha} \, Z_{nl}''(t)],$  (3)

where, $k_1 = \frac{\Gamma[\beta]}{\Gamma[\beta-\alpha+1]}$, with $0 < \alpha \leq 1, 0 < \beta \leq 1$

III. $D^\alpha D^\beta t^m = D^{\alpha+\beta} t^m$ for function derivative of $Z(t) = t^m$, $m \in R^+$
IV. $D^{GFD}(XY) = X D^{GFD}(Y) + Y D^{GFD}(X)$ where $X, Y$ be $\alpha$-differentiable function
V. $D^{GFD}\left(\frac{X}{Y}\right) = \frac{Y D^{GFD}(X) - X D^{GFD}(Y)}{Y^2}$ where $X, Y$ be $\alpha$-differentiable function
VI. $D^\alpha I_\alpha Z(t) = Z(t)$ for $\geq 0$ and $Z$ is any continuous function in the domain.

### 2.1. The Generalized Fractional Nikiforov-Uvarov (NU) Method.

By using generalized fractional derivative, the parametric generalized fractional Nikiforov-Uvarov (NU) method is introduced. The second-order parametric generalized differential equation is exactly solved in the fractional form as in Ref. [27]

$$D^\alpha[D^\alpha \psi(s)] + \frac{\bar{\tau}(s)}{\sigma(s)} D^\alpha \psi(s) + \frac{\bar{\sigma}(s)}{\sigma^2} \psi(s) = 0, \quad (4)$$

where $\bar{\sigma}$, $\sigma(s)$ and $\bar{\tau}(s)$ are polynomials of $2\alpha$-th, $2\alpha$-th and $\alpha$-th degree.
where,

$$\pi(s) = \frac{D^\alpha \sigma(s) - \bar{\tau}(s)}{2} \pm \sqrt{\left(\frac{D^\alpha \sigma(s) - \bar{\tau}(s)}{2}\right)^2 - \bar{\sigma}(s) + K \sigma(s)}, \quad (5)$$

and

$$\lambda = K + D^\alpha \pi(s), \quad (6)$$

$\lambda$ is constant and $\pi(s)$ is $\alpha$-th degree polynomial. The values of $K$ in the square-root of Eq. (5) is possible to determine whether the expression under the square root is square of expression. Replacing $K$ into Eq. (5), we define

$$\tau(s) = \bar{\tau}(s) + 2 \pi(s), \quad (7)$$

the derivative of $\tau$ should be negative [28], since $\rho(s) > 0$ and $\sigma(s) > 0$ then this is solution. If $\lambda$ in Eq. (6) is

$$\lambda = \lambda_n = -n \, D^\alpha \tau - \frac{n(n-1)}{2} D^\alpha[D^\alpha \sigma(s)]. \quad (8)$$

The hypergeometric type equation has a particular solution with degree $\alpha$. Eq. (4) has a solution which is the product of two independent parts

$$\psi(s) = \phi(s)\, y(s), \tag{9}$$

where,

$$y_n(s) = \frac{B_n}{\rho(s)} (D^\alpha)^n (\sigma(s)^n \rho_n(s)), \tag{10}$$

$$D^\alpha[\sigma(s)\, \rho(s)] = \tau(s)\, \sigma(s), \tag{11}$$

$$\frac{D^\alpha \phi(s)}{\phi(s)} = \frac{\pi(s)}{\sigma(s)} \tag{12}$$

### 2.2. Second Order Parametric Generalized Differential Equation

The following equation is a general form of the Schrödinger equation which can be obtained by transforming into a second-order parametric generalized differential equation.

$$D^\alpha[D^\alpha \psi(s)] + \frac{\alpha_1 - \alpha_2 s^\alpha}{s^\alpha(1-\alpha_3 s^\alpha)} D^\alpha \psi(s) + \frac{-\xi_1 s^{2\alpha} + \xi_2 s^\alpha - \xi_3}{(s^\alpha(1-\alpha_3 s^\alpha))^2} \psi(s) = 0. \tag{13}$$

$$\bar{\tau}(s) = \alpha_1 - \alpha_2 s^\alpha, \tag{14}$$

$$\sigma(s) = s^\alpha(1 - \alpha_3 s^\alpha), \tag{15}$$

$$\breve{\sigma}(s) = -\xi_1 s^{2\alpha} + \xi_2 s^\alpha - \xi_3. \tag{16}$$

Substituting these into Eq. (5), we obtain

$$\pi = \alpha_4 + \alpha_5 s^\alpha \pm \sqrt{(\alpha_6 - K\alpha_3) s^{2\alpha} + (\alpha_7 + K) s^\alpha + \alpha_8}, \tag{17}$$

where,

$$\alpha_4 = \tfrac{1}{2}(k_1 \alpha - \alpha_1) \tag{18}$$

$$\alpha_5 = \tfrac{1}{2}(\alpha_2 - 2\alpha_3 k_1 \alpha) \tag{19}$$

$$\alpha_6 = \alpha_5^2 + \xi_1 \tag{20}$$

$$\alpha_7 = 2\alpha_4 \alpha_5 - \xi_2 \tag{21}$$

$$\alpha_8 = \alpha_4^2 + \xi_3 \tag{22}$$

In Eq. (17), the function under square root must be the square of a polynomial according to the NU method, so that

$$K = -(\alpha_7 + 2\alpha_3 \alpha_8) \pm 2\sqrt{\alpha_8 \alpha_9}, \tag{23}$$

where,

$$\alpha_9 = \alpha_3 \alpha_7 + \alpha_3^2 \alpha_8 + \alpha_6. \tag{24}$$

In case $K$ is negative in the form

$$K = -(\alpha_7 + 2\alpha_3 \alpha_8) - 2\sqrt{\alpha_8 \alpha_9} \tag{25}$$

So that $\pi$ becomes

$$\pi = \alpha_4 + \alpha_5 s^\alpha - [(\sqrt{\alpha_9} + \alpha_3 \sqrt{\alpha_8}) s^\alpha - \sqrt{\alpha_8}] \tag{26}$$

From Eqs. (7), (17) and (26), we get

$$\tau = \alpha_1 + 2\alpha_4 - (\alpha_2 - 2\alpha_5) s^\alpha - [(\sqrt{\alpha_9} + \alpha_3 \sqrt{\alpha_8}) s^\alpha - \sqrt{\alpha_8}] \tag{27}$$

From Eqs. (2) and (27), we get,

$$D^\alpha \tau = k_1 [-\alpha(\alpha_2 - 2\alpha_5) - 2\alpha(\sqrt{\alpha_9} + \alpha_3 \sqrt{\alpha_8})] = k_1 [-2\alpha^2 \alpha_3 - 2\alpha(\sqrt{\alpha_9} + \alpha_3 \sqrt{\alpha_8})] < 0 \tag{28}$$

From Eqs. (6, 8), we get the equation of the energy spectrum

$$n k_1 \alpha \alpha_2 - (2n+1) k_1 \alpha \alpha_5 + (2n+1) k_1 \alpha (\sqrt{\alpha_9} + \alpha_3 \sqrt{\alpha_8}) + n(n-1) k_1^2 \alpha^2 \alpha_3 + \alpha_7 + 2\alpha_3 \alpha_8 + 2\sqrt{\alpha_8 \alpha_9} = 0 \tag{29}$$

If $\alpha = 1 = \beta$ then $k_1 = 1$, we get the classical equation of the energy eigenvalue as Ref. [25]

$$n \alpha_2 - (2n+1)\alpha_5 + (2n+1)(\sqrt{\alpha_9} + \alpha_3 \sqrt{\alpha_8}) + n(n-1)\alpha_3 + \alpha_7 + 2\alpha_3 \alpha_8 + 2\sqrt{\alpha_8 \alpha_9} = 0. \tag{30}$$

from Eq. (11), we get

$$\rho(s) = s^{\frac{\alpha_{10}-\alpha}{k_1}} (1 - \alpha_3 s^\alpha)^{\frac{\alpha_{11}}{\alpha k_1 \alpha_3} - \frac{\alpha_{10}}{\alpha k_1} - \frac{1}{k_1}} \quad (31)$$

From Eq. (10), we get

$$y_n = P_n^{(\frac{\alpha_{10}-\alpha}{k_1}, \frac{\alpha_{11}}{\alpha k_1 \alpha_3} - \frac{\alpha_{10}}{\alpha k_1} - \frac{1}{k_1})} (1 - 2\alpha_3 s^\alpha) \quad (32)$$

where,

$$\alpha_{10} = \alpha_1 + 2\alpha_4 + 2\sqrt{\alpha_8} \quad (33)$$

$$\alpha_{11} = \alpha_2 - 2\alpha_5 + 2(\sqrt{\alpha_9} + \alpha_3\sqrt{\alpha_8}) \quad (34)$$

From Eq. (9), the generalized solution of the wave function becomes,

$$\psi(s) = s^{\frac{\alpha_{12}}{k_1}} (1 - \alpha_3 s^\alpha)^{\frac{-\alpha_{13}}{\alpha k_1 \alpha_3} - \frac{\alpha_{12}}{\alpha k_1}} P_n^{(\frac{\alpha_{10}-\alpha}{k_1}, \frac{\alpha_{11}}{\alpha k_1 \alpha_3} - \frac{\alpha_{10}}{\alpha k_1} - \frac{1}{k_1})} (1 - 2\alpha_3 s^\alpha) \quad (35)$$

where,
$P_n^{(\gamma,\delta)}$ are Jacobi polynomials.

$$\alpha_{12} = \alpha_4 + \sqrt{\alpha_8} \quad (36)$$

$$\alpha_{13} = \alpha_5 - (\sqrt{\alpha_9} + \alpha_3\sqrt{\alpha_8}) \quad (37)$$

Some problems, in case $\alpha_3 = 0$.

$$\lim_{\alpha_3 \to 0} P_n^{(\frac{\alpha_{10}-\alpha}{k_1}, \frac{\alpha_{11}}{\alpha k_1 \alpha_3} - \frac{\alpha_{10}}{\alpha k_1} - \frac{1}{k_1})} (1 - \alpha_3 s^\alpha) = L_n^{\frac{\alpha_{10}-\alpha}{k_1}} (\frac{\alpha_{11}}{\alpha k_1} s^\alpha) \quad (38)$$

$$\lim_{\alpha_3 \to 0} (1 - \alpha_3 s^\alpha)^{\frac{-\alpha_{13}}{\alpha k_1 \alpha_3} - \frac{\alpha_{12}}{\alpha k_1}} = e^{\frac{\alpha_{13}}{\alpha k_1} s^\alpha} \quad (39)$$

Eq. (35), becomes

$$\psi(s) = s^{\frac{\alpha_{12}}{k_1}} e^{\frac{\alpha_{13}}{\alpha k_1} s^\alpha} L_n^{\frac{\alpha_{10}-\alpha}{k_1}} (\frac{\alpha_{11}}{\alpha k_1} s^\alpha). \quad (40)$$

Where, $L_n$ being the Laguerre polynomials.
The second solution of Eq. (23) in the following case

$$K = -(\alpha_7 + 2\alpha_3\alpha_8) + 2\sqrt{\alpha_8\alpha_9} \quad (41)$$

then, the wave function is,

$$\psi(s) = s^{\frac{\alpha_{12}^*}{k_1}} (1 - \alpha_3 s^\alpha)^{\frac{-\alpha_{13}^*}{\alpha k_1 \alpha_3} - \frac{\alpha_{12}^*}{\alpha k_1}} P_n^{(\frac{\alpha_{10}^*-\alpha}{k_1}, \frac{\alpha_{11}^*}{\alpha k_1 \alpha_3} - \frac{\alpha_{10}^*}{\alpha k_1} - \frac{1}{k_1})} \times (1 - 2\alpha_3 s^\alpha) \quad (42)$$

The generalized solution of the energy eigenvalue is,

$$nk_1\alpha\alpha_2 - 2nk_1\alpha\alpha_5 + (2n+1)k_1\alpha(\sqrt{\alpha_9} - \alpha_3\sqrt{\alpha_8}) + n(n-1)k_1^2\alpha^2\alpha_3 + \alpha_7 + 2\alpha_3\alpha_8 - 2\sqrt{\alpha_8\alpha_9} + k_1\alpha\alpha_5 = 0 \quad (43)$$

where,

$$\begin{aligned}\alpha_{10}^* &= \alpha_1 + 2\alpha_4 - 2\sqrt{\alpha_8},\\ \alpha_{11}^* &= \alpha_2 - 2\alpha_5 + 2(\sqrt{\alpha_9} - \alpha_3\sqrt{\alpha_8}) \quad \alpha_{12}^* = \alpha_4 - \sqrt{\alpha_8},\\ \alpha_{13}^* &= \alpha_5 - (\sqrt{\alpha_9} - \alpha_3\sqrt{\alpha_8}).\end{aligned} \quad (44)$$

### 3. SOME APPLICATIONS
#### Case (1): Extended Cornell potential

We note that Cornell potential has two features the Coulomb potential and the confinement potential. The Coulomb potential describes the interaction at the short distance and confinement part describes the interaction at the long distances, and the harmonic potential to support the confinement force and it is mainly used to describe bound states of hadrons as in Ref. [29].

$$V(r) = ar^2 + br - \frac{c}{r}, \quad (45)$$

The radial Schrodinger equation where, the interaction potential is the extended Cornell potential defined as in Ref. [29] and $s = e^{-\lambda r}$ that we get,

$$\frac{d^2R}{dr^2} + \frac{1}{s}\frac{dR}{dr} + \frac{1}{s^2(1-s)^2}\{-\xi_1 s^2 + \xi_2 s - \xi_3\}R(s) = 0 \quad (46)$$

where,

$$\xi_1 = -\frac{2\mu E}{\hbar^2 \lambda^2} + t_1, \quad \xi_2 = -\frac{4\mu E}{\hbar^2 \lambda^2} + t_2, \quad \xi_3 = -\frac{2\mu E}{\hbar^2 \lambda^2} + t_3, \quad (47)$$

with,

$$\begin{aligned} t_1 &= \frac{12\mu a}{\lambda^4 \hbar^2} + \frac{6\mu b}{\lambda^3 \hbar^2}, \\ t_2 &= \frac{8\mu a}{\lambda^4 \hbar^2} + \frac{6\mu b}{\lambda^3 \hbar^2} - \frac{2\mu c}{\lambda \hbar^2}, \\ t_3 &= \frac{2\mu a}{\lambda^4 \hbar^2} + \frac{2\mu b}{\lambda^3 \hbar^2} - \frac{2\mu c}{\lambda \hbar^2} + l(l+1), \end{aligned} \quad (48)$$

and we get the generalized fractional radial part of the Schrödinger equation is

$$D^\alpha [D^\alpha R(s)] + \frac{1-s^\alpha}{s^\alpha(1-s^\alpha)} D^\alpha R(s) + \frac{-\xi_1 s^{2\alpha} + \xi_2 s^\alpha - \xi_3}{(s^\alpha(1-s^\alpha))^2} R(s) = 0, \quad (49)$$

By using the following parameters, we get

$$\begin{aligned} \alpha_1 &= 1, \alpha_2 = 1, \alpha_3 = 1, \alpha_4 = \frac{1}{2}(k_1 \alpha - 1), \\ \alpha_5 &= \frac{1}{2}(1 - 2k_1 \alpha), \alpha_6 = \frac{1}{4}(1 - 2k_1 \alpha)^2 - \frac{2\mu E}{\hbar^2 \lambda^2} + t_1, \\ \alpha_7 &= \frac{1}{2}(k_1 \alpha - 1)(1 - 2k_1 \alpha) + \frac{4\mu E}{\hbar^2 \lambda^2} - t_2, \\ \alpha_8 &= \frac{1}{4}(k_1 \alpha - 1)^2 - \frac{2\mu E}{\hbar^2 \lambda^2} + t_3, \alpha_9 = \frac{1}{4}k_1^2 \alpha^2 + t_1 - t_2 + t_3, \\ \alpha_{10} &= k_1 \alpha + 2\sqrt{\frac{1}{4}(k_1 \alpha - 1)^2 - \frac{2\mu E}{\hbar^2 \lambda^2} + t_3}, \\ \alpha_{11} &= 2k_1 \alpha + 2\left(\sqrt{\frac{1}{4}k_1^2 \alpha^2 + t_1 - t_2 + t_3} + \sqrt{\frac{1}{4}(k_1 \alpha - 1)^2 - \frac{2\mu E}{\hbar^2 \lambda^2} + t_3}\right), \\ \alpha_{12} &= \frac{1}{2}(k_1 \alpha - 1) + \sqrt{\frac{1}{4}(k_1 \alpha - 1)^2 - \frac{2\mu E}{\hbar^2 \lambda^2} + t_3}, \\ \alpha_{13} &= \frac{1}{2}(1 - 2k_1 \alpha) - \left(\sqrt{\frac{1}{4}k_1^2 \alpha^2 + t_1 - t_2 + t_3} + \sqrt{\frac{1}{4}(k_1 \alpha - 1)^2 - \frac{2\mu E}{\hbar^2 \lambda^2} + t_3}\right), \end{aligned} \quad (50)$$

We get, the generalized fractional of the energy eigenvalue is

$$E = \frac{\hbar^2 \lambda^2}{2\mu}\left(t_3 + \frac{1}{4}(k_1 \alpha - 1)^2\right) - \frac{\hbar^2 \lambda^2}{2\mu}\left(\frac{t_1 - t_3 - \left(\left(n+\frac{1}{2}\right)k_1 \alpha + \sqrt{\frac{1}{4}k_1^2 \alpha^2 + t_1 - t_2 + t_3}\right)^2}{2\left(\left(n+\frac{1}{2}\right)k_1 \alpha + \sqrt{\frac{1}{4}k_1^2 \alpha^2 + t_1 - t_2 + t_3}\right)}\right)^2 \quad (51)$$

where, $t = t_1 - t_2 + t_3$,

The generalized fractional of the wave function is,

$$\psi(s) = A\, s^{\frac{\frac{1}{2}(k_1 \alpha - 1) + \sqrt{\frac{1}{4}(k_1 \alpha - 1)^2 - \frac{2\mu E}{\hbar^2 \lambda^2} + t_3}}{k_1}} (1 s^\alpha)^{\frac{-\left(\frac{1}{2}(1 - 2k_1 \alpha) - \left(\sqrt{\frac{1}{4}k_1^2 \alpha^2 + t} + \sqrt{\frac{1}{4}(k_1 \alpha - 1)^2 - \frac{2\mu E}{\hbar^2 \lambda^2} + t_3}\right)\right)}{k_1 \alpha}} \cdot \frac{\frac{1}{2}(k_1 \alpha - 1) + \sqrt{\frac{1}{4}(k_1 \alpha - 1)^2 - \frac{2\mu E}{\hbar^2 \lambda^2} + t_3}}{k_1 \alpha}$$

$$P_n^{\left(\frac{-\alpha + k_1 \alpha + 2\sqrt{\frac{1}{4}(k_1 \alpha - 1)^2 - \frac{2\mu E}{\hbar^2 \lambda^2} + t_3}}{k_1},\; \frac{2k_1 \alpha + 2\left(\sqrt{\frac{1}{4}k_1^2 \alpha^2 + t} + \sqrt{\frac{1}{4}(k_1 \alpha - 1)^2 - \frac{2\mu E}{\hbar^2 \lambda^2} + t_3}\right)}{k_1 \alpha} - \frac{k_1 \alpha + 2\sqrt{\frac{1}{4}(k_1 \alpha - 1)^2 - \frac{2\mu E}{\hbar^2 \lambda^2} + t_3}}{k_1 \alpha} - \frac{1}{k_1}\right)} \times (1 - 2 s^\alpha) \quad (52)$$

where A is a normalization constant.

If $\alpha = \beta = 1$ then $k_1 = 1$, so that we get the energy eigenvalue and the corresponding wave function as Ref. [29]. as follows

$$E = \frac{\hbar^2 \lambda^2}{2\mu} t_3 - \frac{\hbar^2 \lambda^2}{2\mu}\left(\frac{t_1 - t_3 - \left(\left(n+\frac{1}{2}\right) + \sqrt{\frac{1}{4} + t_1 - t_2 + t_3}\right)^2}{2\left(\left(n+\frac{1}{2}\right) + \sqrt{\frac{1}{4} + t_1 - t_2 + t_3}\right)}\right)^2 \quad (53)$$

$$\psi(s) = A\, s^{\sqrt{-\frac{2\mu E}{\hbar^2 \lambda^2} + t_3}} (1-s)^{\frac{1}{2} + \sqrt{\frac{1}{4} + t}} P_n^{(2\sqrt{-\frac{2\mu E}{\hbar^2 \lambda^2} + t_3},\, 2\sqrt{\frac{1}{4} + t})} (1 - 2s) \quad (54)$$

**Case (2): Pesudoharmonic Potential**

Pesudoharmonic Potential is mainly used to describe bound states of spectroscopy of die-atomic molecules and may be used for energy spectra of linear and non-linear systems, see Refs. [30,31] for details.

$$V(r) = V_0 \left(\frac{r}{r_0} - \frac{r_0}{r}\right)^2, \tag{55}$$

where, $V_0$ is dissociation energy between two atoms in a solid, $r_0$ is the equilibrium intermolecular separation and $r$ is the internuclear separation.

After transformation $s = A^2 r^2$, where $(A = 1\ (eV))$, we get the generalized fractional radial part of the Schrödinger equation is

$$D^\alpha [D^\alpha R(s)] + \frac{3/2}{s^\alpha} D^\alpha R(s) + \frac{-\gamma^2 s^{2\alpha} + \varepsilon s^\alpha - \beta}{s^{2\alpha}} R(s) = 0. \tag{56}$$

where, the following dimensionless parameters

$$\left.\begin{aligned}
\gamma^2 &= \frac{V_0 \mu}{2 r_0^2 A^4 \hbar^2}, \\
\varepsilon &= \frac{\mu}{\hbar^2 A^2} (E_{n,l} + 2 V_0) \\
\beta &= \frac{\mu}{\hbar^2} \left(V_0 r_0^2 + \frac{L(L+1)\hbar^2}{2\mu}\right),
\end{aligned}\right\} \tag{57}$$

by using the following parameters,

$$\left.\begin{aligned}
\xi_1 &= \gamma^2, \xi_2 = \varepsilon, \xi_3 = \beta, \\
\alpha_1 &= \frac{3}{2}, \alpha_2 = 0, \alpha_3 = 0, \alpha_4 = \frac{1}{2}(k_1 \alpha - 3/2), \\
\alpha_5 &= 0, \alpha_6 = \gamma^2, \alpha_7 = -\varepsilon, \alpha_8 = \frac{1}{4}(k_1 \alpha - 3/2)^2 + \beta, \\
\alpha_9 &= \gamma^2, \alpha_{10} = k_1 \alpha + 2\sqrt{\frac{1}{4}(k_1 \alpha - 3/2)^2 + \beta}, \alpha_{11} = 2\gamma, \\
\alpha_{12} &= \frac{1}{2}(k_1 \alpha - 3/2) + \sqrt{\frac{1}{4}(k_1 \alpha - 3/2)^2 + \beta}, \alpha_{13} = -\gamma,
\end{aligned}\right\} \tag{58}$$

We get, the generalized fractional of the energy eigenvalue is,

$$\varepsilon = (2n+1)\gamma k_1 \alpha + 2\sqrt{\frac{1}{4}(k_1 \alpha - 3/2)^2 + \beta}, \tag{59}$$

And, the generalized fractional of the wave function is,

$$\left.\begin{aligned}
\psi(s) &= s^{\frac{1}{k_1}\left(\frac{1}{2}\left(k_1 \alpha - \frac{3}{2}\right) + \sqrt{\frac{1}{4}\left(k_1 \alpha - \frac{3}{2}\right)^2 + \beta}\right)} e^{\frac{-\gamma}{k_1 \alpha} s^\alpha} \\
& L_n^{\frac{1}{k_1}\left(k_1 \alpha + 2\sqrt{\frac{1}{4}\left(k_1 \alpha - \frac{3}{2}\right)^2 + \beta}\right) - \alpha} \left(\frac{2\gamma}{k_1 \alpha} s^\alpha\right),
\end{aligned}\right\} \tag{60}$$

If $\alpha = \beta = 1$ then $k_1 = 1$, so that we get the energy eigenvalue and the corresponding wave function as Ref. [25].

$$\varepsilon = \left((2n+1) + 2\sqrt{\frac{1}{16} + \beta}\right)\gamma, \tag{61}$$

$$\psi(s) = s^{\left(\frac{-1}{4} + \sqrt{\frac{1}{16} + \beta}\right)} e^{-\gamma s} L_n^{\left(2\sqrt{\frac{1}{16} + \beta}\right)} (2\gamma s), \tag{62}$$

**Case (3): Mie Potential**

The Mie potential possess the general features of the true interaction energy and it is special kind of exactly solvable power-law and inverse power-law potentials other than the Coulombic and harmonic oscillator as in Refs. [32-33]

$$V(r) = V_0 \left(\frac{1}{2}\left(\frac{a}{r}\right)^2 - \frac{a}{r}\right) \tag{63}$$

where, $V_0$ is dissociation energy, $a$ is the positive constant which is strongly repulsive at shorter distances and $r$ is the internuclear separation

After transforming $s = r\,A$, $(A = 1eV)$, we get the generalized fractional radial part of the Schrödinger equation is

$$D^\alpha[\,D^\alpha R(s)] + \frac{2}{s^\alpha} D^\alpha R(s) + \frac{\varepsilon^2 s^{2\alpha} - \beta s^\alpha - \gamma}{s^{2\alpha}} R(s) = 0. \tag{64}$$

where, the following dimensionless parameters:

$$\left.\begin{array}{l}\varepsilon^2 = \dfrac{2\mu}{\hbar^2 A^2} E_{n,l}, \\ \beta = -\dfrac{2\mu}{\hbar^2 A} V_0 a, \\ \gamma = \dfrac{2\mu}{\hbar^2}\left(\dfrac{1}{2}V_0 a^2 + \dfrac{L(L+1)\hbar^2}{2\mu}\right).\end{array}\right\} \tag{65}$$

By using the following parameters,

$$\left.\begin{array}{l}\xi_1 = -\varepsilon^2,\ \xi_2 = -\beta,\ \xi_3 = \gamma, \\ \alpha_1 = 2,\ \alpha_2 = 0,\ \alpha_3 = 0,\ \alpha_4 = \dfrac{1}{2}(k_1\alpha - 2), \\ \alpha_5 = 0,\ \alpha_6 = -\varepsilon^2,\ \alpha_7 = \beta,\ \alpha_8 = \dfrac{1}{4}(k_1\alpha - 2)^2 + \gamma \\ \alpha_9 = -\varepsilon^2,\ \alpha_{10} = k_1\alpha + 2\sqrt{\dfrac{1}{4}(k_1\alpha - 3/2)^2 + \gamma}, \\ \alpha_{11} = 2\sqrt{-\varepsilon^2},\ \alpha_{12} = \dfrac{1}{2}(k_1\alpha - 2) + \sqrt{\dfrac{1}{4}(k_1\alpha - 2)^2 + \gamma}, \\ \alpha_{13} = -\sqrt{-\varepsilon^2}.\end{array}\right\} \tag{66}$$

The generalized fractional of the energy eigenvalue is given,

$$-\varepsilon^2 = \beta^2\left((2n+1)k_1\alpha + 2\sqrt{\tfrac{1}{4}(k_1\alpha - 2)^2 + \gamma}\right)^{-2}, \tag{67}$$

The generalized fractional of the wave function is,

$$\psi(s) = A\, s^{\frac{1}{k_1}\left(\frac{1}{2}(k_1\alpha - 2) + \sqrt{\frac{1}{4}(k_1\alpha - 2)^2 + \gamma}\right)} e^{\frac{-i\varepsilon}{k_1\alpha} s^\alpha} L_n^{\frac{1}{k_1}\left(k_1\alpha + 2\sqrt{\frac{1}{4}(k_1\alpha - 2)^2 + \gamma}\right) - \alpha}\left(\tfrac{2\,i\,\varepsilon}{k_1\alpha} s^\alpha\right), \tag{68}$$

If $\alpha = \beta = 1$ then $k_1 = 1$, so that we get the energy eigenvalue and the corresponding wave function as Ref. [25].

$$\varepsilon^2 = -\beta^2\left(2n + 1 + \sqrt{1 + 4\gamma}\right)^{-2} \tag{69}$$

$$\psi(s) = A\, s^{\frac{-1}{2} + \frac{1}{2}\sqrt{1+4\gamma}}\, e^{-i\varepsilon s} L_n^{\sqrt{1+4\gamma}}(2\,i\,\varepsilon\, s), \tag{70}$$

### Case (4): Kratzer-Fues Potential

The Kratzer-Fues potential has a long-range attraction and a repulsive part and it is approaches infinity as the inter-nuclear distance approaches zero and it is used to describe molecular structure between two atoms as in Ref. [34]

$$V(r) = D_e\left(\frac{r - r_e}{r}\right)^2 \tag{71}$$

After transforming $s = r\,A$, $A = 1eV$ the generalized fractional radial part of the Schrödinger equation is

$$D^\alpha[\,D^\alpha R(s)] + \frac{2}{s^\alpha} D^\alpha R(s) + \frac{\varepsilon^2 s^{2\alpha} - \beta s^\alpha - \gamma}{s^{2\alpha}} R(s) = 0, \tag{72}$$

where, the following dimensionless parameters

$$\left.\begin{array}{l}\varepsilon^2 = \dfrac{2\mu(E_n - D_e)}{A^2 \hbar^2}, \\ \beta = \dfrac{-4\,\mu\, D_e\, r_e}{A\,\hbar^2}, \\ \gamma = \dfrac{2\,\mu\left(D_e\, r_e^2 + \dfrac{l(l+1)\hbar^2}{2\mu}\right)}{\hbar^2},\end{array}\right\} \tag{73}$$

By using the following parameters,

$$\begin{aligned}
\xi_1 &= -\varepsilon^2, \xi_2 = -\beta, \xi_3 = \gamma, \\
\alpha_1 &= 2, \alpha_2 = 0, \alpha_3 = 0, \alpha_4 = \frac{1}{2}(k_1\alpha - 2), \\
\alpha_5 &= 0, \alpha_6 = -\varepsilon^2, \alpha_7 = \beta, \alpha_8 = \frac{1}{4}(k_1\alpha - 2)^2 + \gamma \\
\alpha_9 &= -\varepsilon^2, \alpha_{10} = k_1\alpha + 2\sqrt{\frac{1}{4}(k_1\alpha - 2)^2 + \gamma}, \\
\alpha_{11} &= 2\sqrt{-\varepsilon^2}, \alpha_{12} = \frac{1}{2}(k_1\alpha - 2) + \sqrt{\frac{1}{4}(k_1\alpha - 2)^2 + \gamma}, \\
\alpha_{13} &= -\sqrt{-\varepsilon^2},
\end{aligned}$$  (74)

the generalized fractional solution of eigenvalue is given,

$$-\varepsilon^2 = \beta^2 \left((2n+1)k_1\alpha + 2\sqrt{\frac{1}{4}(k_1\alpha - 2)^2 + \gamma}\right)^{-2},$$ (75)

The generalized fractional of the wave function is,

$$\psi(s) = A\, s^{\frac{1}{k_1}\left(\frac{1}{2}(k_1\alpha - 2) + \sqrt{\frac{1}{4}(k_1\alpha - 2)^2 + \gamma}\right)} e^{\frac{-i\varepsilon}{k_1\alpha}s^\alpha} L_n^{\frac{1}{k_1}\left(k_1\alpha + 2\sqrt{\frac{1}{4}(k_1\alpha - 2)^2 + \gamma}\right) - \alpha}\left(\frac{2i\varepsilon}{k_1\alpha}s^\alpha\right),$$ (76)

If $\alpha = \beta = 1$ then $k_1 = 1$, so that we get the energy eigenvalue and the corresponding wave function as Ref. [25]

$$-\varepsilon^2 = \beta^2(2n + 1 + \sqrt{1 + 4\gamma})^{-2}$$ (77)

$$\psi(s) = A\, s^{-\frac{1}{2} + \frac{1}{2}\sqrt{1+4\gamma}}\, e^{-i\varepsilon s} L_n^{\sqrt{1+4\gamma}}(2i\varepsilon s),$$ (78)

**Case (5): Harmonic Oscillator Potential**

Harmonic Oscillator Potential function is as in Ref. [35]

$$V(r) = \frac{1}{2} m\omega^2 r^2,$$ (79)

where, $\omega$ is the angular frequency of the oscillator.
The generalized fractional radial part of the Schrödinger equation is

$$D^\alpha[D^\alpha\psi(s)] + \frac{1}{2s^\alpha}\psi^\alpha R(s) + \frac{-s^{2\alpha} + \beta^2 s^\alpha - l(l+1)}{4s^{2\alpha}}\psi(s) = 0.$$ (80)

where,

$$\beta^2 = \frac{2E}{\hbar\omega}$$ (81)

By using the following parameters,

$$\begin{aligned}
\xi_1 &= \frac{1}{4}, \xi_2 = \frac{1}{4}\beta^2, \xi_3 = \frac{1}{4}l(l+1), \\
\alpha_1 &= \frac{1}{2}, \alpha_2 = 0, \alpha_3 = 0, \alpha_4 = \frac{1}{2}(k_1\alpha - \frac{1}{2}), \\
\alpha_5 &= 0, \alpha_6 = \frac{1}{4}, \alpha_7 = \frac{-1}{4}\beta^2, \alpha_8 = \frac{1}{4}(k_1\alpha - \frac{1}{2})^2 + \frac{1}{4}l(l+1), \\
\alpha_9 &= \frac{1}{4}, \alpha_{10} = k_1\alpha + \sqrt{(k_1\alpha - 1/2)^2 + l(l+1)}, \alpha_{11} = 1, \\
\alpha_{12} &= \frac{1}{2}(k_1\alpha - \frac{1}{2}) + \frac{1}{2}\sqrt{(k_1\alpha - \frac{1}{2})^2 + l(l+1)}, \alpha_{12} = \frac{-1}{2},
\end{aligned}$$  (82)

the generalized fractional eigenvalue is given,

$$E = \hbar\omega\left[(2n+1)k_1\alpha + \sqrt{(k_1\alpha - 1/2)^2 + l(l+1)}\right]$$ (83)

The generalized fractional wave function is

$$\psi(s) = s^{\frac{1}{k_1}\left((k_1\alpha)+\sqrt{\left(k_1\alpha-\frac{1}{2}\right)^2+l(l+1)}\right)} e^{\frac{-1}{k_1\alpha}s^\alpha} \times L_n^{\frac{1}{k_1}\left(k_1\alpha+1/2\sqrt{(k_1\alpha-1/2)^2+l(l+1)}\right)-\frac{\alpha}{k_1}}\left(\frac{1}{k_1\alpha}s^\alpha\right) \quad (84)$$

If $\alpha = \beta = 1$ then $k_1 = 1$, so that we get the energy eigenvalue and the corresponding wave function as Ref. [35]

$$E = \hbar\omega\left[(2n+1)+\sqrt{\frac{1}{4}+l(l+1)}\right] \quad (85)$$

$$\psi(s) = s^{\left(1+\sqrt{\frac{1}{4}+l(l+1)}\right)} e^{-s} L_n^{\left(1/2\sqrt{\frac{1}{4}+l(l+1)}\right)}(s). \quad (86)$$

### Case (6): Morse Potential

The Morse potential has contributed a significant role in describing the interaction of atoms in diatomic and polyatomic molecules as in Refs. [36, 37]

$$V(r) = D_0(1-e^{-\delta r})^2 \quad (87)$$

where $D_0$ is dissociation energy and $\delta$ is the range of the potential. After transforming the generalized fractional radial part of the Schrödinger equation is

$$D^\alpha[D^\alpha\psi(s)] + \frac{1}{s^\alpha}D^\alpha\psi(s) + \frac{-Ps^{2\alpha}+Qs^\alpha-R}{s^{2\alpha}}\psi(s) = 0. \quad (88)$$

where,

$$\varepsilon_n^2 = \frac{-2\mu E}{\hbar^2}, \gamma = \frac{2\mu D_0}{\hbar^2}, P = \frac{\gamma}{\delta^2}, Q = \frac{2\gamma}{\delta^2}, R = \frac{\varepsilon_n^2+\gamma}{\delta^2}, \quad (89)$$

By using the following parameters,

$$\left.\begin{aligned}
\xi_1 &= P, \xi_2 = Q, \xi_3 = R, \\
\alpha_1 &= 1, \alpha_2 = 0, \alpha_3 = 0, \alpha_4 = \frac{1}{2}(k_1\alpha-1), \\
\alpha_5 &= 0, \alpha_6 = P, \alpha_7 = -Q, \alpha_8 = \frac{1}{4}(k_1\alpha-1)^2 + R \\
\alpha_9 &= P, \alpha_{10} = k_1\alpha + 2\sqrt{\frac{1}{4}(k_1\alpha-1)^2 + R}, \alpha_{11} = 2\sqrt{P}, \\
\alpha_{12} &= \frac{1}{2}(k_1\alpha-1) + \sqrt{\frac{1}{4}(k_1\alpha-1)^2+R}, \alpha_{13} = -\sqrt{P},
\end{aligned}\right\} \quad (90)$$

the generalized fractional of the energy eigenvalue is given,

$$E = D_0 - \frac{\hbar^2\delta^2}{8\mu}\left[-(k_1\alpha-1)^2 + \left((2n+1)k_1\alpha - 2\sqrt{\frac{2\mu D_0}{\hbar^2\delta^2}}\right)^2\right] \quad (91)$$

The generalized fractional of the wave function is

$$\psi(s) = \mathbb{N}_n s^{\frac{\frac{1}{2}(k_1\alpha-1)+\sqrt{\frac{1}{4}(k_1\alpha-1)^2+R}}{k_1}} e^{\left(\frac{-\sqrt{P}}{k_1\alpha}\right)s^\alpha} L_n^{\frac{k_1\alpha+2\sqrt{\frac{1}{4}(k_1\alpha-1)^2+R}-\alpha}{k_1}}\left(\frac{2\sqrt{P}}{k_1\alpha}s^\alpha\right) \quad (92)$$

where, $\mathbb{N}_n$ is a normalization constant,

If $\alpha = \beta = 1$ then $k_1 = 1$, so that we get the energy eigenvalue and the corresponding wave function as in Ref. [36].

$$E = D_0 - \frac{\hbar^2\delta^2}{8\mu}\left[\left((2n+1) - 2\sqrt{\frac{2\mu D_0}{\hbar^2\delta^2}}\right)^2\right] \quad (93)$$

$$\psi(s) = \mathbb{N}_n s^{\sqrt{R}} e^{(-\sqrt{P})s} L_n^{2\sqrt{R}}(2\sqrt{P}s) \quad (94)$$

### Case (7): Woods-Saxon Potential

It is used to describe heavy-ion reactions which the interaction of a neutron with a heavy nucleus as in Refs. [26, 38-39]

$$V(r) = \frac{-V_0}{1+e^{\frac{r-R_0}{a}}} \quad (95)$$

where $V_0$ is the potential depth, $R_0$ is width of the potential, $r - R_0 \equiv r$ and $\frac{1}{a} = 2\lambda$

After transforming the generalized fractional radial part of the Schrödinger equation is

$$D^\alpha [D^\alpha \psi(s)] + \frac{1 - q s^\alpha}{s^\alpha (1 - q s^\alpha)} D^\alpha \psi(s) + \frac{-\varepsilon q^2 s^{2\alpha} + (2 \varepsilon q - \beta q) s^\alpha + \beta - \varepsilon}{(s^\alpha(1 - q s^\alpha))^2} \psi(s) = 0. \quad (96)$$

where,

$$\varepsilon = \frac{-mE}{2\hbar^2 \lambda^2} > 0, \qquad \beta = \frac{mV_0}{2\hbar^2 \lambda^2}, \quad (97)$$

By using the following parameters,

$$\begin{aligned}
\xi_1 &= \varepsilon q^2, \xi_2 = 2\varepsilon q - \beta q, \xi_3 = \varepsilon - \beta, \\
\alpha_1 &= 1, \alpha_2 = q, \alpha_3 = q, \alpha_4 = \frac{1}{2}(k_1 \alpha - 1), \\
\alpha_5 &= \frac{1}{2} q (1 - 2k_1 \alpha), \alpha_6 = \frac{1}{4} q^2 (1 - 2k_1 \alpha)^2 + \varepsilon q^2, \\
\alpha_7 &= \frac{1}{2} q (1 - 2k_1 \alpha)(k_1 \alpha - 1) - 2\varepsilon q - \beta q, \\
\alpha_8 &= \frac{1}{4}(k_1 \alpha - 1)^2 + \varepsilon - \beta, \alpha_9 = \frac{1}{4} q^2 k_1^2 \alpha^2, \\
\alpha_{10} &= k_1 \alpha + 2\sqrt{\frac{1}{4}(k_1 \alpha - 1)^2 + \varepsilon - \beta}, \\
\alpha_{11} &= 3 k_1 \alpha q + 2 q \sqrt{\frac{1}{4}(k_1 \alpha - 1)^2 + \varepsilon - \beta}, \\
\alpha_{12} &= \frac{1}{2}(k_1 \alpha - 1) + \sqrt{\frac{1}{4}(k_1 \alpha - 1)^2 + \varepsilon - \beta}, \\
\alpha_{13} &= \frac{1}{2} q (1 - 2k_1 \alpha) - \left(\frac{1}{2} k_1 \alpha q + q \sqrt{\frac{1}{4}(k_1 \alpha - 1)^2 + \varepsilon - \beta}\right).
\end{aligned} \right\} \quad (98)$$

The generalized fractional of the energy eigenvalue is given,

$$\varepsilon = -\frac{1}{4}(k_1 \alpha - 1)^2 + \frac{\beta}{2} + \frac{\beta^2}{(2(n+1)k_1 \alpha)^2} + \left(\frac{(n+1)k_1 \alpha}{2}\right)^2, \quad (99)$$

The generalized fractional of the wave function is,

$$\psi(s) = \mathbb{N}_n \, s^{\frac{1}{k_1}\left(\frac{1}{2}(k_1 \alpha - 1) + \sqrt{\frac{1}{4}(k_1 \alpha - 1)^2 + \varepsilon - \beta}\right)} (1 - q s^\alpha)^{\left(\frac{-\left(\frac{1}{2}q(1-2k_1\alpha) - \left(\frac{1}{2}k_1 \alpha q + q\sqrt{\frac{1}{4}(k_1\alpha - 1)^2 + \varepsilon - \beta}\right)\right)}{k_1 \alpha q} - \frac{\frac{1}{2}(k_1\alpha - 1) + \sqrt{\frac{1}{4}(k_1\alpha-1)^2 + \varepsilon - \beta}}{k_1 \alpha}\right)} \times$$
$$P_n^{\left(\frac{-\alpha + k_1 \alpha + 2\sqrt{\frac{1}{4}(k_1\alpha-1)^2 + \varepsilon - \beta}}{k_1}, \frac{3 k_1 \alpha q + 2 q \sqrt{\frac{1}{4}(k_1\alpha-1)^2 + \varepsilon-\beta}}{k_1 \alpha q} - \frac{k_1 \alpha + 2\sqrt{\frac{1}{4}(k_1\alpha-1)^2 + \varepsilon - \beta}}{k_1 \alpha} - \frac{1}{k_1}\right)} \times (1 - 2 q s^\alpha) \quad (100)$$

If $\alpha = \beta = 1$ then $k_1 = 1$, and $q = 1$, so that we get the energy eigenvalue and the corresponding wave function as in Ref. [26]

$$\varepsilon = \frac{\beta}{2} + \frac{\beta^2}{(2(n+1))^2} + \left(\frac{(n+1)}{2}\right)^2, \quad (101)$$

$$\psi(s) = \mathbb{N}_n \, s^{\sqrt{\varepsilon - \beta}} (1 - s) P_n^{(2\sqrt{\varepsilon - \beta}, 1)}(1 - 2s) \quad (102)$$

**Case (8): Hulthen Potential**

The Hulthen Potential is a short-range potential and it is obtained in the form as in Ref. [40]

$$V(r) = -\frac{P}{p} \frac{1}{e^{\frac{r}{p}} - 1}, \quad (103)$$

where, $P, p$ are the strength and the range parameter of the potential function.

After transforming the generalized fractional radial part of the Schrödinger equation is

$$D^\alpha[D^\alpha\psi(s)] + \frac{1-s^\alpha}{s^\alpha(1-s^\alpha)}D^\alpha\psi(s) + \frac{-(A+B)s^{2\alpha}+(2A+B-C)s^\alpha-A}{(s^\alpha(1-s^\alpha))^2}\psi(s) = 0. \qquad (104)$$

where,

$$A = \frac{-2\mu E p^2}{\hbar^2}, B = \frac{2\mu E p P}{\hbar^2}, C = l(l+1). \qquad (105)$$

By using the following parameters,

$$\begin{aligned}
\xi_1 &= A+B, \xi_2 = (2A+B-C), \xi_3 = A, \\
\alpha_1 &= 1, \alpha_2 = 1, \alpha_3 = 1, \alpha_4 = \frac{1}{2}(k_1\alpha - 1), \\
\alpha_5 &= \frac{1}{2}(1 - 2k_1\alpha), \alpha_6 = \frac{1}{4}(1 - 2k_1\alpha)^2 + A + B, \\
\alpha_7 &= \frac{1}{2}(k_1\alpha - 1)(1 - 2k_1\alpha) - 2A - B + C \; \alpha_8 = \frac{1}{4}(k_1\alpha - 1)^2 + A, \\
\alpha_9 &= \frac{1}{4}k_1^2\alpha^2 + C, \alpha_{10} = k_1\alpha + 2\sqrt{\frac{1}{4}(k_1\alpha - 1)^2 + A}, \\
\alpha_{11} &= 2k_1\alpha + 2(\sqrt{\frac{1}{4}k_1^2\alpha^2 + C} + \sqrt{\frac{1}{4}(k_1\alpha - 1)^2 + A}), \\
\alpha_{12} &= \frac{1}{2}(k_1\alpha - 1) + \sqrt{\frac{1}{4}(k_1\alpha - 1)^2 + A}, \\
\alpha_{13} &= \frac{1}{2}(1 - 2k_1\alpha) - (\sqrt{\frac{1}{4}k_1^2\alpha^2 + C} + \sqrt{\frac{1}{4}(k_1\alpha - 1)^2 + A}).
\end{aligned} \qquad (106)$$

The generalized fractional of the energy eigenvalue is given

$$E_n = \frac{-\hbar^2}{8\mu p^2}\left\{\frac{\left[\frac{2\mu p P}{\hbar^2} - l(l+1) - \frac{k_1\alpha}{2}\left(k_1\alpha + \sqrt{k_1^2\alpha^2 + 4l(l+1)}\right) - nk_1\alpha\left(\frac{(n+1)}{2}k_1\alpha + \sqrt{k_1^2\alpha^2 + 4l(l+1)}\right)\right]^2 - (k_1\alpha-1)^2}{\left[nk_1\alpha + \frac{1}{2}(k_1\alpha + \sqrt{k_1^2\alpha^2 + 4l(l+1)})\right]^2}\right\} \qquad (107)$$

The generalized fractional of the wave function is

$$\psi(s) = \mathbb{N}_n s^{\frac{1}{k_1}\left(\frac{1}{2}(k_1\alpha-1)+\sqrt{\frac{1}{4}(k_1\alpha-1)^2+A}\right)}(1-s^\alpha)^{\frac{-\left(\frac{1}{2}(1-2k_1\alpha)-(\sqrt{\frac{1}{4}k_1^2\alpha^2+C}+\sqrt{\frac{1}{4}(k_1\alpha-1)^2+A})\right)}{k_1\alpha}-\frac{\frac{1}{2}(k_1\alpha-1)+\sqrt{\frac{1}{4}(k_1\alpha-1)^2+A}}{k_1\alpha}}$$
$$P_n^{\frac{\left(k_1\alpha+2\sqrt{\frac{1}{4}(k_1\alpha-1)^2+A}\right)-\alpha}{k_1},\frac{2k_1\alpha+2(\sqrt{\frac{1}{4}k_1^2\alpha^2+C}+\sqrt{\frac{1}{4}(k_1\alpha-1)^2+A})}{k_1\alpha}-\frac{k_1\alpha+2\sqrt{\frac{1}{4}(k_1\alpha-1)^2+A}}{k_1\alpha}-\frac{1}{k_1}}(1-2s^\alpha) \qquad (108)$$

If $\alpha = \beta = 1$ then $k_1 = 1$, so that we get the energy eigenvalue and the corresponding wave function as in Ref. [26]

$$E_n = \frac{-\hbar^2}{8\mu p^2}\left\{\frac{\left[\frac{2\mu p P}{\hbar^2}-l(l+1)-\frac{1}{2}(1+\sqrt{1+4l(l+1)})-n\left(\frac{(n+1)}{2}+\sqrt{1+4l(l+1)}\right)\right]^2}{\left[n+\frac{1}{2}(1+\sqrt{1+4l(l+1)})\right]^2}\right\} \qquad (109)$$

$$\psi(s) = \mathbb{N}_n s^{\sqrt{A}}(1-s)^{-\left(\frac{-1}{2}-(\sqrt{\frac{1}{4}+C}+\sqrt{A})\right)-\sqrt{A}}P_n^{2\sqrt{A},2(\sqrt{C})}(1-2s) \qquad (110)$$

### Case (9): Deformed Rosen-Morse Potential

Deformed Rosen-Morse potential is used to describe bound state of die-atomic molecules and it is given in the following form as in Ref. [41]

$$V(r) = \frac{V_1}{1+q e^{-2\alpha^* r}} - V_2 q \frac{e^{-2\alpha^* r}}{(1+q e^{-2\alpha^* r})^2} \qquad (111)$$

After transforming $-s = e^{-2\alpha^* r}$ the generalized fractional radial part of the Schrödinger equation is

$$D^\alpha[D^\alpha\psi(s)] + \frac{1-q s^\alpha}{s^\alpha(1-q s^\alpha)}D^\alpha\psi(s) + \frac{-\varepsilon q^2 s^{2\alpha}+(2\varepsilon q + \beta q - \gamma)s^\alpha - (\beta+\varepsilon)}{(s^\alpha(1-q s^\alpha))^2}\psi(s) = 0. \qquad (112)$$

where,

$$\varepsilon = \frac{-\mu E}{2\alpha^{*2}\hbar^2}, \beta = \frac{\mu V_1 q}{2\alpha^{*2}\hbar^2}, \gamma = \frac{\mu V_2 q}{2\alpha^{*2}\hbar^2}, \tag{113}$$

By using the following parameters,

$$\begin{aligned}
\xi_1 &= \varepsilon q^2, \xi_2 = (2\varepsilon q + \beta q - \gamma), \xi_3 = (\beta + \varepsilon),\\
\alpha_1 &= 1, \alpha_2 = q, \alpha_3 = q, \alpha_4 = \frac{1}{2}(k_1\alpha - 1),\\
\alpha_5 &= \frac{q}{2}(1 - 2k_1\alpha), \alpha_6 = \frac{q^2}{4}(1 - 2k_1\alpha)^2 + \varepsilon q^2,\\
\alpha_7 &= \frac{q}{2}(k_1\alpha - 1)(1 - 2k_1\alpha) - 2\varepsilon q - \beta q + \gamma\\
\alpha_8 &= \frac{1}{4}(k_1\alpha - 1)^2 + \beta + \varepsilon,\\
\alpha_9 &= \frac{q^2}{4}k_1^2\alpha^2 + \gamma q, \alpha_{10} = k_1\alpha + 2\sqrt{\frac{1}{4}(k_1\alpha - 1)^2 + \beta + \varepsilon},\\
\alpha_{11} &= 2k_1q\alpha + 2\left(\sqrt{\frac{1}{4}k_1^2q^2\alpha^2 + \gamma q} + q\sqrt{\frac{1}{4}(k_1\alpha - 1)^2 + \beta + \varepsilon}\right),\\
\alpha_{12} &= \frac{1}{2}(k_1\alpha - 1) + \sqrt{\frac{1}{4}(k_1\alpha - 1)^2 + \beta + \varepsilon},\\
\alpha_{13} &= \frac{q}{2}(1 - 2k_1\alpha) - \left(\sqrt{\frac{1}{4}k_1^2q^2\alpha^2 + \gamma q} + q\sqrt{\frac{1}{4}(k_1\alpha - 1)^2 + \beta + \varepsilon}\right),
\end{aligned} \tag{144}$$

we get, the generalized fractional of the energy eigenvalue is,

$$\varepsilon = \frac{-1}{4}(k_1\alpha - 1)^2 - \frac{\beta}{2} + \frac{\beta^2}{\left((2n+1)k_1\alpha + \sqrt{k_1^2\alpha^2 + \frac{4\gamma}{q}}\right)^2} + \frac{1}{16}\left((2n+1)k_1\alpha + \sqrt{k_1^2\alpha^2 + \frac{4\gamma}{q}}\right)^2 \tag{115}$$

The generalized fractional of the wave function is,

$$\psi(s) = s^{\frac{\frac{1}{2}(k_1\alpha - 1) + \sqrt{\frac{1}{4}(k_1\alpha - 1)^2 + \beta + \varepsilon}}{k_1}} (1 - qs^\alpha)^{\frac{-\left(\frac{q}{2}(1-2k_1\alpha) - \left(\sqrt{\frac{1}{4}k_1^2q^2\alpha^2 + \gamma q} + q\sqrt{\frac{1}{4}(k_1\alpha-1)^2 + \beta + \varepsilon}\right)\right)}{k_1\alpha q} - \frac{\frac{1}{2}(k_1\alpha - 1) + q\sqrt{\frac{1}{4}(k_1\alpha-1)^2 + \beta + \varepsilon}}{k_1\alpha}}$$
$$P_n^{\left(\frac{k_1\alpha + 2\sqrt{\frac{1}{4}(k_1\alpha-1)^2 + \beta + \varepsilon} - \alpha}{k_1}, \frac{2k_1q\alpha + 2\left(\sqrt{\frac{1}{4}k_1^2\alpha^2 + \gamma q} + q\sqrt{\frac{1}{4}(k_1\alpha-1)^2 + \beta + \varepsilon}\right)}{k_1\alpha q} - \frac{k_1\alpha + 2\sqrt{\frac{1}{4}(k_1\alpha-1)^2 + \beta + \varepsilon}}{k_1\alpha} - \frac{1}{k_1}\right)} (1 - 2qs^\alpha) \tag{116}$$

If $\alpha = \beta = 1$ then $k_1 = 1$, so that we get the energy eigenvalue and the corresponding wave function as in Ref. [25]

$$\varepsilon = -\frac{\beta}{2} + \frac{\beta^2}{\left((2n+1) + \sqrt{1 + \frac{4\gamma}{q}}\right)^2} + \frac{1}{16}\left((2n+1) + \sqrt{1 + \frac{4\gamma}{q}}\right)^2, \tag{117}$$

$$\psi(s) = s^{\sqrt{\beta + \varepsilon}} (1 - qs)^{\frac{1}{2}\left(1 + \sqrt{1 + \frac{4\gamma}{q}}\right)} P_n^{(2\sqrt{\beta + \varepsilon}, \sqrt{1 + \frac{4\gamma}{q}})} (1 - 2qs) \tag{118}$$

### Case (10): Pöschl-Teller Potential

Pöschl-Teller Potential used to describe bound state of die-atomic molecules and it is given in the following form as in Refs. [42-43]

$$V(r) = -4V_0 \frac{e^{-2\alpha^* r}}{(1 + qe^{-2\alpha^* r})^2} \tag{119}$$

After transforming $-s = e^{-2\alpha^* r}$, the generalized fractional radial part of the Schrödinger equation is

$$D^\alpha[D^\alpha\psi(s)] + \frac{1 - qs^\alpha}{s^\alpha(1 - qs^\alpha)} D^\alpha\psi(s) + \frac{-\varepsilon^2 q^2 s^{2\alpha} + (2\varepsilon^2 q - \beta^2)s^\alpha - \varepsilon^2}{(s^\alpha(1 - qs^\alpha))^2} \psi(s) = 0. \tag{120}$$

where,

$$\varepsilon^2 = \frac{-\mu E}{2\alpha^{*2}\hbar^2}, \beta^2 = \frac{2\mu V_0}{\alpha^{*2}\hbar^2} \qquad (121)$$

By using the following parameters,

$$\begin{aligned}
\xi_1 &= \varepsilon^2 q^2, \xi_2 = (2\varepsilon^2 q - \beta^2), \xi_3 = \varepsilon^2, \\
\alpha_1 &= 1, \alpha_2 = q, \alpha_3 = q, \alpha_4 = \frac{1}{2}(k_1\alpha - 1), \\
\alpha_5 &= \frac{q}{2}(1 - 2k_1\alpha), \alpha_6 = \frac{q^2}{4}(1 - 2k_1\alpha)^2 + \varepsilon^2 q^2, \\
\alpha_7 &= \frac{q}{2}(k_1\alpha - 1)(1 - 2k_1\alpha) - 2\varepsilon^2 q + \beta^2 \; \alpha_8 = \frac{1}{4}(k_1\alpha - 1)^2 + \varepsilon^2, \\
\alpha_9 &= \frac{q^2}{4}k_1^2\alpha^2 + \beta^2 q, \alpha_{10} = k_1\alpha + 2\sqrt{\frac{1}{4}(k_1\alpha - 1)^2 + \varepsilon^2}, \\
\alpha_{11} &= 2qk_1\alpha + 2\left(\sqrt{\frac{1}{4}k_1^2 q^2\alpha^2 + \beta^2 q} + q\sqrt{\frac{1}{4}(k_1\alpha - 1)^2 + \varepsilon^2}\right) \\
\alpha_{12} &= \frac{1}{2}(k_1\alpha - 1) + \sqrt{\frac{1}{4}(k_1\alpha - 1)^2 + \varepsilon^2}, \\
\alpha_{13} &= \frac{q}{2}(1 - 2k_1\alpha) - \left(\sqrt{\frac{1}{4}k_1^2 q^2\alpha^2 + \beta^2 q} + q\sqrt{\frac{1}{4}(k_1\alpha - 1)^2 + \varepsilon^2}\right),
\end{aligned} \qquad (122)$$

the generalized fractional of the energy eigenvalue is given,

$$\varepsilon = \frac{-1}{4}(k_1\alpha - 1)^2 - \frac{1}{4}\left((2n+1)k_1\alpha + \sqrt{k_1^2\alpha^2 + \frac{4\beta^2}{q}}\right). \qquad (123)$$

The generalized fractional of the wave function is given,

$$\psi(s) = s^{\frac{\frac{1}{2}(k_1\alpha - 1) + \sqrt{\frac{1}{4}(k_1\alpha - 1)^2 + \varepsilon^2}}{k_1}} (1 - qs^\alpha)^{\frac{-\left(\frac{q}{2}(1 - 2k_1\alpha) - \left(\sqrt{\frac{1}{4}k_1^2 q^2\alpha^2 + \beta^2 q} + q\sqrt{\frac{1}{4}(k_1\alpha - 1)^2 + \varepsilon^2}\right)\right)}{k_1\alpha q} - \frac{\frac{1}{2}(k_1\alpha - 1) + \sqrt{\frac{1}{4}(k_1\alpha - 1)^2 + \varepsilon^2}}{k_1\alpha}}$$

$$P_n^{\left(\frac{k_1\alpha + 2\sqrt{\frac{1}{4}(k_1\alpha - 1)^2 + \varepsilon^2}}{k_1} - \alpha, \frac{2k_1 q\alpha + 2\left(\sqrt{\frac{1}{4}k_1^2\alpha^2 + \beta^2 q} + q\sqrt{\frac{1}{4}(k_1\alpha - 1)^2 + \varepsilon^2}\right)}{k_1\alpha q} - \frac{k_1\alpha + 2\sqrt{\frac{1}{4}(k_1\alpha - 1)^2 + \varepsilon^2}}{k_1\alpha} - \frac{1}{k_1}\right)}(1 - 2qs^\alpha) \qquad (124)$$

If $\alpha = \beta = 1$ then $k_1 = 1$, so that we get the energy eigenvalue and the corresponding wave function as in Ref. [25].

$$\varepsilon = -\frac{1}{4}\left((2n+1) + \sqrt{1 + \frac{4\beta^2}{q}}\right), \qquad (125)$$

$$\psi(s) = s^\varepsilon (1 - qs)^{\frac{1}{2}\left(1 + \sqrt{1 + \frac{4\beta^2}{q}}\right)} P_n^{\left(2\varepsilon, \sqrt{1 + \frac{4\beta^2}{q}}\right)}(1 - 2qs) \qquad (126)$$

### 4. CONCLUSION

By using generalized fractional derivative, we obtained the solution the parametric second-order differential equation by using the NU method which is more effective than the power series method, numerical methods, or approximation methods because we get on good results by using this method as in Refs. [22,23]. The parametric second-order differential equation is the general case and we get the special case when $\alpha = \beta = 1$ as in Ref. [25]. We get a solution of the Schrödinger equation by using the parametric generalized fractional Nikiforov-Uvarov (NU) method and we get the energy eigenvalues and the corresponding wave function for some known potentials such that the Cornell potential, the pesudoharmonic potential, the Mie Potential, the Kratzer-Fues potential, the harmonic oscillator potential, the Morse potential, the Woods-Saxon potential, the Hulthen Potential, the deformed Rosen-Morse Potential and the Pöschl-Teller Potential. At $\alpha = \beta = 1$, we get the special classical solutions of Refs. [25,26,29,35,36]. These applications play an important role in the fields of molecular and hadronic physics.

### ORCID

**M. Abu-Shady**, https://orcid.org/0000-0001-7077-7884

**ПАРАМЕТРИЧНИЙ УЗАГАЛЬНЕНИЙ ДРОБОВИЙ МЕТОД НІКІФОРОВА-УВАРОВА ТА ЙОГО ЗАСТОСУВАННЯ**
**М. Абу-Шаді[a], Х.М. Фатх-Аллах[b]**

[a]*Кафедра математики та комп'ютерних наук, факультет природничих наук, Університет Менуфія, Єгипет*
[b]*Вищий інженерно-технологічний інститут, Менуфія, Єгипет*

За допомогою узагальненої дробової похідної введено параметричний узагальнений дробовий метод Нікіфорова-Уварова (НУ). Параметричне узагальнене диференціальне рівняння другого порядку точно розв'язується у дробовій формі. Отримані результати застосовано до розширеного потенціалу Корнелла, псевдогармонійного потенціалу, потенціалу Мі, потенціалу Кратцера-Фюса, потенціалу гармонічного осцилятора, потенціалу Морзе, потенціалу Вудса-Саксона, потенціалу Хюльтена, деформованого потенціалу Розена-Морса. і потенціалу Пошля-Теллера, які відіграють важливу роль у галузях молекулярної та атомної фізики. Особливі класичні випадки отримані з дробових випадків $α = β = 1$, які узгоджуються з останніми роботами.
**Ключові слова:** *нерелятивістські моделі; узагальнена дробова похідна; молекулярна фізика; адронна фізика*